\documentclass[12pt]{article}

\ifx\pdfoutput\undefined
\usepackage[dvips,bookmarks]{hyperref}
\else
\usepackage{hyperref}
\fi
\hypersetup{colorlinks=false,bookmarksopen,bookmarksnumbered,citecolor=blue,
   pdfstartview=FitH}

\usepackage[dvips]{graphicx}
\usepackage{latexsym}

\usepackage{amssymb,amsfonts,amsmath}
\usepackage{graphicx} 
\usepackage{indentfirst}

 \usepackage{bbm}

\parskip=\medskipamount

\newcommand {\cD}{{\cal D}}

\newcommand {\cF}{{\cal F}}

\newcommand {\cH}{{\cal H}}

\newcommand {\cK}{{\cal K}}
\newcommand {\cL}{{\cal L}}
\newcommand {\cM}{{\cal M}}
\newcommand {\cN}{{\cal N}}
\newcommand {\cO}{{\cal O}}

\newcommand {\cU}{{\cal U}}

\def\a{\alpha}
\def \bi{\bibitem}

\def\b{\beta}
\def\c{\chi}
\def\d{\delta}
\def\e{\epsilon}
\def\f{\phi}

\def\G{\Gamma}

\def\l{\lambda}
\def\m{\mu}
\def\n{\nu}
\def\o{\omega}

\def\q{\theta}
\def\r{\rho}
\def\s{\sigma}
\def\t{\tau}
\def\u{\upsilon}
\def\x{\xi}
\def\z{\zeta}

\def\F{\Phi}
\def\J{\Psi}
\def\L{\Lambda}
\def\O{\Omega}

\def\S{\Sigma}
\def\U{\Upsilon}
\def\X{\Xi}

\def\rd{{\rm d}}

\newcommand{\ad}{{\dot{\alpha}}}                           %new
\newcommand{\ve}{\varepsilon}                            %new
\newcommand{\pa}{\partial}                           %new
\newcommand{\hf}{\frac12}
\newcommand{\vf}{\varphi}
\newcommand{\be}{\begin{equation}}
\newcommand{\ee}{\end{equation}}
\newcommand{\bea}{\begin{eqnarray}}
\newcommand{\eea}{\end{eqnarray}}
\newcommand{\non}{\nonumber}
\newcommand{\1}{\underline{1}}
\newcommand{\2}{\underline{2}}

\def\dt#1{{\buildrel {\hbox{\LARGE .}} \over {#1}}}    % dot-over for sp/sb

\newcommand{\bm}[1]{\mbox{\boldmath$#1$}}

\def\double #1{#1{\hbox{\kern-2pt $#1$}}}

\begin{document}
%%%%%%%%%%%%%%%%
%%%%%%%%%%%%%%%%
\begin{titlepage}
\begin{flushright}
CQUeST-2009-0298\\
October, 2009 \\
\end{flushright}
\vspace{5mm}

\begin{center}
{\Large \bf  \mbox{$\bm{ \cN=2}$}  supersymmetric sigma-models
and duality}\footnote{Based in part on  lectures given 
at the Center for Quantum Spacetime,
Sogang University, Seoul, October 2009.} 
\\ 
\end{center}

\begin{center}

{\bf
Sergei M. Kuzenko\footnote{kuzenko@cyllene.uwa.edu.au}
} \\
\vspace{5mm}

\footnotesize{
{\it School of Physics M013, The University of Western Australia\\
35 Stirling Highway, Crawley W.A. 6009, Australia}}  
~\\
\vspace{2mm}

\end{center}
\vspace{5mm}

\begin{abstract}
\baselineskip=14pt
${}$For two families of four-dimensional off-shell $\cN=2$ supersymmetric 
nonlinear $\s$-models constructed originally in projective superspace, 
we develop their formulation in terms of $\cN=1 $ chiral superfields.
Specifically, these theories are: (i) $\s$-models on cotangent bundles  $T^*\cM$ 
of arbitrary real analytic K\"ahler manifolds $\cM$; 
(ii) general superconformal $\s$-models described by weight-one 
polar supermultiplets. Using superspace techniques, 
we obtain a universal expression for the holomorphic symplectic two-form
${\bm \o}^{(2,0)}$ which determines the second supersymmetry transformation
and is associated with the two complex structures of 
the hyperk\"ahler space  $T^*\cM$ that are complimentary 
to the one induced from $\cM$. This two-form is shown to coincide with the canonical
 holomorphic symplectic structure. In the case (ii),  we demonstrate that ${\bm \o}^{(2,0)}$
and the homothetic conformal Killing vector  
determine the explicit form of  the superconformal transformations.
At the heart of our construction is the duality (generalized Legendre transform)
between off-shell $\cN=2$ supersymmetric nonlinear $\s$-models 
and their on-shell $\cN=1$ chiral realizations.
We finally present the most general $\cN=2$ superconformal nonlinear $\s$-model
formulated in terms of $\cN=1 $ chiral superfields.
The approach developed can naturally be generalized in order to 
describe 5D and 6D  superconformal nonlinear $\s$-models in 4D $\cN=1$ superspace.
\end{abstract}
\vspace{1cm}

\vfill
\end{titlepage}

\newpage
\renewcommand{\thefootnote}{\arabic{footnote}}
\setcounter{footnote}{0}

\tableofcontents{}
\vspace{1cm}
\bigskip\hrule

%%%%%%%%%%%%%%%%%%%%%%%%%%%%%%%%%%%%%%%%%%%%%%%%%%
%%%%%%%%%%%%%%%%%%%%%%%%%%%%%%%%%%%%%%%%%%%%%%%%%%

\section{Introduction}
\setcounter{equation}{0}

Four-dimensional $\cN=2$ supersymmetric nonlinear $\s$-models
can be formulated in terms of component fields  \cite{A-GF,BW,DeJdeWKV} or 
$\cN=1$ chiral superfields \cite{LR,HKLR,BX}.
These constructions are quite elegant and geometric, especially the one in $\cN=1$
superspace. However, they both present just the existence theorems in the sense that 
their practical usefulness  is extremely  limited if one is interested
in the explicit construction of general $\cN=2$ supersymmetric nonlinear $\s$-models
(or, equivalently, hyperk\"ahler metrics).
Achieving such a goal requires the use of $\cN=2$ superspace techniques, 
and the appropriate setting turns out to be the so-called projective superspace 
approach
\cite{KLR,LR-projective} (see also \cite{LR2008} for a recent review).\footnote{Projective 
superspace can be derived from harmonic superspace \cite{harm1,harm2} in a singular limit
\cite{K98,JS}. However, the two approaches are truly complementary. While 
 the harmonic formalism is indispensable for quantum calculations in $\cN=2$ super Yang-Mills
 theories, the projective formalism is ideal for $\s$-model constructions.
 It should be remarked that both approaches make use of the isotwistor superspace
${\mathbb R}^{4|8}\times {\mathbb C}P^1$  pioneered by Rosly 
\cite{Rosly}. }
The power of this approach in the context of $\cN=2$ supersymmetric nonlinear $\s$-models
is due to the following reasons: 

(i) the $\cN=2$ supersymmetric action is generated by a Lagrangian that can be chosen to be 
an arbitrary function (modulo some mild restrictions) of several superfield dynamical
variables -- off-shell $\cN=2$ projective supermultiplets; 

(ii)  such supermultiplets  are naturally decomposed into a set of standard $\cN=1$ 
superfields. \\
The latter property in fact implies an intimate connection between
the projective superspace approach and the $\cN=1$ superspace construction of \cite{LR,HKLR,BX}.
It is pertinent here to elaborate on this point in some more detail. For simplicity, 
our discussion will be restricted to the case of a single hypermultiplet.

In the projective superspace setting, there are infinitely many off-shell realizations 
for a {\it neutral hypermultiplet}, each of which is characterized by a finite number of auxiliary fields. 
Such off-shell supermultiplets are labelled by a positive integer $n=2,3, \dots$, 
and are called real $\cO(2n)$ multiplets.\footnote{The case $n=1$ corresponds to the $\cN=2$ 
tensor multiplet \cite{Wess}. This multiplet  is very special, because (i) it involves no 
auxiliary superfields $\cU_\imath$; and (ii) the physical linear superfield becomes real,
$\S =\bar \S$.}
${}$For a {\it charged hypermultiplet}, 
there exists a single off-shell realization with an infinite number of auxiliary fields,
which is called the polar hypermultiplet (the terminology follows \cite{G-RRWLvU}). 
All of these $\cN=2$ multiplets can readily be decomposed 
into a set of standard $\cN=1$ superfields, and the corresponding content is the following:
two {\it physical} superfields $\F$ and $\S$ and their conjugates $\bar \F$ and $\bar \S$, 
as well as some number of {\it auxiliary} superfields 
$\cU_\imath$, where the index $\imath$ may take a finite ($2n-3$ in the case of $\cO(2n)$ multiplets)
or infinite number of values (for the polar multiplet).
The physical superfields $\F$ and $\S$ are chiral and complex linear, respectively,
\be
{\bar D}_{\dt{\a}} \F =0~, \qquad \qquad {\bar D}^2 \S = 0 ~,
\label{chiral+linear}
\ee
while the auxiliary superfields $\cU_\imath$ are unconstrained.
Upon reduction to  $\cN=1$ superspace, 
the action functional of an  $\cN=2$ supersymmetric $\s$-model  
takes the form
\bea
S =  \int  \rd^4 x\,{\rm d}^4\q\, 
L_{\rm off-shell}(\F , \bar \F, \S , \bar \S, \cU_\imath)~,
\label{af1}
\eea
for some Lagrangian $L_{\rm off-shell}(\F , \bar \F, \S , \bar \S, \cU_\imath)$.
Although this action is formulated in $\cN=1$ superspace, and thus  
only its $\cN=1$ supersymmetry is manifest, it is in fact invariant under off-shell
$\cN=2$ supersymmetry transformations provided (\ref{af1}) 
is derived from a manifestly $\cN=2$ supersymmetric  action in projective superspace 
\cite{LR-projective}.
The superfields $\cU_\imath$ are auxiliary because they are unconstrained and appear 
in the Lagrangian without derivatives. They can be integrated out, at least in principle,
using the corresponding equations 
of motion 
\bea
\frac{\pa }{\pa \cU_{\jmath} }L_{\rm off-shell}(\F , \bar \F, \S , \bar \S, \cU_\imath)=0
\quad \Longrightarrow \quad 
\cU_\imath = \cU_\imath (\F , \bar \F, \S , \bar \S)~.
\eea
As a result, one arrives at an action formulated in terms of the physical superfields 
only,
\bea
S =  \int  \rd^4 x\,{\rm d}^4\q\, 
L_{\rm on-shell}(\F , \bar \F, \S , \bar \S)~.
\label{af2}
\eea
This action is of course $\cN=2$ supersymmetric; however 
the corresponding  transformations form a closed algebra on the mass shell only.
Since $\S$ is complex linear, the formulation (\ref{af2}) cannot be used directly 
to link the theory under consideration with the results of \cite{LR,HKLR,BX}.
In order to obtain a formulation in terms of chiral superfields  only, 
one has to dualize  $\S$ and $\bar \S$ into a chiral superfield and 
its conjugate.\footnote{The existence of duality between the chiral and the complex linear
superfields was noticed for the first time by  Zumino \cite{Zumino1980}
(see also \cite{GS}). This observation naturally extended the duality 
 between the chiral and the real linear superfields \cite{Siegel}.
General aspects of duality in nonlinear $\s$-models in $\cN=1$ superspace 
were elaborated in \cite{LR}.} 
The action (\ref{af2}) is equivalent to the following first-order action:
\bea
S_{\rm first-order}=   \int \rd^4 x\,{\rm d}^4\q \, 
\Big\{\,
L_{\rm on-shell}(\F , \bar \F, \S , \bar \S)~
+\J \,\S + {\bar \J} {\bar \S} 
\Big\}~.
\eea
Here  $\S$ is complex unconstrained, 
while $\Psi$ is chiral, ${\bar D}_{\dt \a} \J =0$.
Integrating out $\S$ and $\bar \S$ leads to an action of the form
\bea
S_{\rm dual}=   \int \rd^4 x\,{\rm d}^4\q \, 
H(\F , \bar \F, \J , \bar \J)~.
\eea
By construction, this theory is $\cN=2$ supersymmetric. It is 
formulated in terms of $\cN=1$ chiral superfields, 
in the spirit of  \cite{LR,HKLR,BX},
and therefore the  Lagrangian $H$ is the hyperk\"ahler potential 
of the target space. What we have described here is known as 
the generalized Legendre transform procedure formalized by Lindstr\"om and Ro\v{c}ek 
twenty years ago \cite{LR-projective}. In the most interesting case of polar 
hypermultiplet self-couplings, nontrivial examples of the generalized Legendre transform
have been given over the last ten years \cite{K98,GK1,GK2,AN,AKL1,AKL2,KN}.

Presently, an interesting open problem is to formulate general $\cN=2$ superconformal 
nonlinear $\s$-models in terms of $\cN=1$ chiral superfields. 
Its significance  follows from several fundamental results in supersymmetry
and their implications. Quaternion K\"ahler manifolds are 
of special importance for supersymmetric theories with eight supercharges, 
for they present target spaces of matter hypermultiplets in $\cN=2$ supergravity \cite{BW}. 
There exists a one-to-one correspondence  \cite{Swann,Galicki} 
between $4n$-dimensional quaternion K\"ahler manifolds
and $4(n+1)$-dimensional hyperk\"ahler spaces possessing a homothetic conformal
Killing vector, and hence an isometric action of SU(2) rotating the complex structures \cite{GR}. 
Such hyperk\"ahler spaces,  known  as ``hyperk\"ahler cones''  in the physics literature, 
turn out to be the target spaces for rigid $\cN=2$ superconformal 
$\s$-models \cite{deWKV0,deWKV,deWRV}.
The quaternion K\"ahler manifolds emerge  as the $\cN=2$ superconformal quotient of the 
corresponding hyperk\"ahler cones \cite{deWRV}. 
The quotient construction (including the K\"ahler reduction from the hyperk\"ahler cone to 
the twistor space of  the quaternion K\"ahler manifold)
can naturally be carried out \cite{deWRV} if the $\cN=2$ superconformal $\s$-model
is realized in terms of $\cN=1$ chiral superfields.
The above consideration indicates 
that the problem of generating 
arbitrary quaternion K\"ahler metrics is essentially 
equivalent to the following two problems in rigid supersymmetry:  (i) construction
of general off-shell $\cN=2$ superconformal $\s$-models in projective superspace;
and (ii) their on-shell re-formulation in terms of $\cN=1$ chiral superfields.

${}$Four-dimensional off-shell $\cN=2$ superconformal multiplets in  projective suprspace
and their couplings 
were described in detail in \cite{K-hyper}, building on the earlier equivalent  results in five dimensions
\cite{K-compactified}. In particular, the most general $\s$-model couplings 
of superconformal weight-one polar hypermultiplets were given in \cite{K-hyper}. First 
steps toward developing the chiral formulation in $\cN=1$ superspace for the $\s$-models given 
were also undertaken in \cite{K-hyper}. The analysis was based on the idea that the 
$\cN=2$ superconformal $\s$-models of weight-one polar hypermultiplets
form a subclass in the family of the off-shell $\cN=2$ supersymmetric $\s$-models
on cotangent bundles of K\"ahler manifolds \cite{K98,GK1,GK2}.
In the present paper, we complete the chiral formulation in $\cN=1$ superspace 
for the superconformal $\s$-models 
introduced in \cite{K-hyper}.

This paper is organized as follows. Section 2 provides a review of the formulation
for $\cN=2 $ supersymmetric nonlinear $\s$-models in terms of $\cN=1$ superfields,
which was pioneered in \cite{HKLR}. In section 3 we develop the chiral formulation in $\cN=1$ superspace 
for general off-shell $\cN=2$ supersymmetric $\s$-models
on cotangent bundles of K\"ahler manifolds. This analysis is extended in section 4
to the case of general off-shell $\cN=2$ superconformal 
$\s$-models of weight-one polar hypermultiplets.
In section 5 we propose a chiral formulation in $\cN=1$ superspace
for the most general $\cN=2$ superconformal nonlinear $\s$-model.
A brief discussion of the results obtained and their possible extensions 
is given in section 6. In the appendix, we provide a purely superspace proof
of the conditions given in \cite{HKLR} for the $\s$-model (\ref{N=1sigma-model}) 
to be invariant under the transformations (\ref{LR-ansatz}) and (\ref{parameter}).
This appendix makes the present paper essentially self-contained.

\section{N = 2 supersymmetric nonlinear sigma-models in N = 1 superspace}
\setcounter{equation}{0}

In this section we review the formulation
for $\cN=2 $ supersymmetric nonlinear $\s$-models in terms of $\cN=1$ superfields,
which was given in \cite{HKLR}.

We start with a general $\cN=1$ supersymmetric nonlinear $\s$-model
\cite{Zumino} 
\bea
S&=& \int {\rm d}^4 x \,{\rm d}^4 \q  \, K\big(\f^a, {\bar \f}^{\overline{b}}\big)~, 
\qquad {\bar D}_{\dt \a} \f^a =0~, 
\label{N=1sigma-model}
\eea
with $K$  the K\"ahler potential of a K\"ahler manifold $\cM$, 
and look for those restrictions on the target space geometry which 
make the theory  be $\cN=2$ supersymmetric.

To describe the second supersymmetry, one makes  \cite{LR,HKLR}
the  ansatz:\footnote{In the linear $\s$-model case, 
the ansatz (\ref{LR-ansatz}) has its origin in $\cN=2$ superspace. 
One should start by considering the Fayet-Sohnius hypermultiplet \cite{Fayet,Sohnius} 
described by a $\cN=2$ isospinor superfield $q^i$ obeying the constraints
$\cD^{(i}_\a q^{j)} = {\bar \cD}^{(i}_{\dt \a} q^{j)} =0$. This multiplet and its
supersymmetry transformation law can readily be reduced to $\cN=1$ superspace;
in particular, $q^i$ generates two $\cN=1$ multiplets $\f_+$ and ${\bar \f}_-$, 
with $\f^a:=( \f_+, \f_-)  $ chiral superfields. For the second supersymmetry transformation, 
one gets eq. (\ref{LR-ansatz}) in which ${\bar \O}^\pm = \pm{\bar \f}^\mp$.
This procedure was carried out explicitly in \cite{GIO} and implicitly in \cite{RT}.}
\bea
\d \f^a &=& \hf {\bar D}^2 \big( {\bar \e} \,{\bar \O}^a \big)~, \qquad 
\d {\bar \f}^{\overline a}  = \hf D^2 \big( { \e} \,\O^{\overline a} \big) ~,
\label{LR-ansatz}
\eea 
for some functions $\O^{\overline a} =\O^{\overline a} \big(\f, \bar \f \big)$
associated with  the K\"ahler manifold $\cM$. The transformation parameter
$\e$ is constrained by 
\be
{\bar D}_{\dt \a} \e =  \pa_{\a \dt \a} \e =  D^2\e =0~\quad \Longleftrightarrow \quad 
\e = \e (\q) = \t + \e^\a \q_\a 
~, \quad \t ={\rm const}~, \quad 
\e^\a ={\rm const}
\label{parameter}
\ee
Here  $\e^\a $ is the supersymmetry parameter,
while $\t$  corresponds to  a central charge transformation.
If the action is invariant under the second supersymmetry transformation
described by $\e_\a$, 
then the central charge symmetry is generated by commuting the first 
({\it manifestly realized}) and the second supersymmetry transformations.

The action is invariant under the central charge transformation provided
\bea
{\bar \o}_{ \bar{b}  \bar{c} } :=  g_{ \bar{b} a}\, {\bar \O}^a{}_{,\bar{c} }
 =-  {\bar \o}_{ \overline{c}  \bar{b} }~, \qquad
  {\bar \O}^a{}_{,\bar{c} }:= \pa_{\bar c}  {\bar \O}^a~,
\label{se1}
\eea
with $g_{a \bar{b} } =K_{a\bar b}:= \pa_a \pa_{\bar b} K$ the K\"ahler metric.
The action is invariant under the 
transformation generated by the parameter $\e_\a$  
if the two-form\footnote{It will be explained shortly why
 $\o_{bc}$ has to be a globally defined two-form on $\cM$.}
  $\o_{bc}$ and its conjugate
${\bar \o}_{ \overline{b}  \overline{c} } $ are covariantly constant,
\begin{subequations}
\bea
\nabla_a  {\bar \o}_{ \overline{b}  \overline{c}  }
&=& \pa_a {\bar \o}_{ \overline{b}  \overline{c}  }=0 ~, 
\label{se2}\\
\nabla_{\bar a}  {\bar \o}_{ \overline{b}  \overline{c}  }&=&0 ~.
\label{se2b}
\eea
\end{subequations}

On the mass shell, 
\bea
 {\bar D}^2 K_a=0 ~,~
\label{mass-shell}
\eea
 and the first and the second supersymmetry transformations
generate the $\cN=2$ super-Poincar\'e algebra {\it without} central charge
provided
\bea
{\bar \O}^a{}_{, \bar c}  \, { \O}^{\bar c}{}_{,  b}
=- \d^a{}_b~.
\label{se3}
\eea
In fact, the closure of the supersymmetry algebra
requires two more conditions
\bea
 {\bar D}^2 {\bar \O}^a&=&0 ~,\\
{\bar \O}^d{}_{, \bar b}  \nabla_d {\bar \O}^a{}_{, \bar c}
- {\bar \O}^d{}_{, \bar c}  \nabla_d {\bar \O}^a{}_{, \bar b} &=&0 ~.
\eea
They hold due to (\ref{se2}) -- (\ref{se3}).

On the mass shell, 
the supersymmetry transformation  (\ref{LR-ansatz})
takes the form:
\bea
\d \f^a = {\bar \e}_{\dt \a}  \,{\bar \O}^a{}_{,\bar b} \,{\bar D}^{\dt \a} {\bar \f}^{\bar b}~.
\eea 
Since $\d \f^a$ should be a vector field on $\cM$, we conclude that ${\bar \O}^a{}_{,\bar b} $
is a tensor field on $\cM$, and therefore $\o_{ab} $ is a two-form.

Let $J \equiv J_3 $ be the complex structure chosen on the target space $\cM$,
\bea
J_3 = \left(
\begin{array}{cc}
{\rm i} \, \d^a{}_b  ~ & ~ 0 \\
0 ~ &   -{\rm i} \, \d^{\bar a}{}_{\bar b}  
\end{array}
\right)~.
\label{J3}
\eea
The above consideration shows that there are two more complex structures defined as
\bea
J_1 = \left(
\begin{array}{cc}
0  ~ & ~ {\bar \O}^a{}_{, \bar b} \\
{\O}^{\bar a}{}_{,  b} ~ &   0
\end{array}
\right)~, \qquad 
J_2 = \left(
\begin{array}{cc}
0  ~ &  {\rm i}\,  {\bar \O}^a{}_{, \bar b} \\
-{\rm i}\, {\O}^{\bar a}{}_{,  b} ~ &   0
\end{array}
\right)~
\label{J1J2}
\eea
such that $\cM$ is K\"ahler with respect to all of them, and 
the operators $J_A = (J_1,J_2,J_3) $ form the quaternionic algebra:
\be
J_A \,J_B = -\d_{AB} \, {\mathbbm 1} + \ve_{ABC}J_C~.
\ee
As a result, it has been demonstrated that the target space $\cM$ is a hyperk\"ahler 
manifold.

As is seen from (\ref{J1J2}), the complex structures are given in terms of 
the tensor fields ${\bar \O}^a{}_{, \bar b}$ and ${\O}^{\bar a}{}_{,  b}$, while the supersymmetry 
transformation (\ref{LR-ansatz}) involves ${\bar \O}^a$ and ${\O}^{\bar a}$. 
The latter can be constructed using the K\"ahler potential \cite{HKLR}:
\be
{\bar \O}^a = \o^{ab}  \big(\f \big) K_b \big(\f , \bar \f \big)~.
\label{barOmega}
\ee
Under the  K\"ahler transformations 
\be
K \big(\f , \bar \f \big)  \quad \longrightarrow \quad K\big(\f , \bar \f \big) 
+  \L \big(\f  \big) + {\bar \L} \big( \bar \f \big)   ~,
\ee
${\bar \O}^a $ changes  as follows:
$ \o^{ab}  K_b \to  \o^{ab}  K_b +  \o^{ab}  \L_b$.
However, the supersymmetry variation
$\d \f^a = \hf {\bar D}^2 \big( {\bar \e} \,{\bar \O}^a \big)$ in (\ref{LR-ansatz}) 
is invariant under the K\"ahler transformations, as emphasized in \cite{BX}.
This completes our review of \cite{HKLR}.

Most of the above relations were given in \cite{HKLR} without proof.
Their {\it purely superspace} proof turns out to be nontrivial and quite interesting 
in its own right. It is described in the Appendix.

It should be pointed out that not all restrictions (\ref{parameter}), 
which were originally put forward in \cite{HKLR},
are necessary. In fact, it is sufficient to restrict the parameter in (\ref{LR-ansatz})
to obey the constraint:
\bea
D_\a \e = {\rm const}~.
\label{parameter2}
\eea
This leaves the following freedom in the choice of  $\e $ in  (\ref{LR-ansatz}): 
\bea
\e = \e (\q) + {\bar \m }~, \qquad D_\a \bar \m =0~,
\eea
with $\e (\q) $ given in (\ref{parameter}).
Choosing $\e  $ in (\ref{LR-ansatz}) to be an antichiral superfield $\bar \m$
provides  an example of {\it trivial} symmetries of the form 
\be
\d \vf^i = \G^{ij} \, \frac{\d S[\vf] }{\d \vf^j}~, \qquad \G^{ij}=-\G^{ji} 
\ee
any theory $S[\vf]$ of bosonic fields $\vf^i$ possesses.
In particular, one can use such a trivial invariance to modify
the second supersymmetry transformation on the manner \cite{BX}:
\bea
\d \f^a &=& \hf {\bar D}^2 \Big( \Big[ \e (\q) +{\bar \e} (\bar \q) \Big]{\bar \O}^a \Big)~, \qquad 
\d {\bar \f}^{\overline a}  = \hf D^2 \Big( \Big[ \e (\q) +{\bar \e} (\bar \q) \Big] {\O^{\overline a}} \Big) ~.
\eea 
This results in {\it no} gain at all in four space-time dimensions.
However, such a form of  supersymmetry transformation is very useful in five and six dimensions.

\section{Non-superconformal nonlinear sigma-models}
\setcounter{equation}{0}

In this section we will investigate four-dimensional  off-shell $\cN=2$
supersymmetric nonlinear $\s$-models that are described in 
ordinary $\cN=1$ 
superspace by the action
\bea
S[\U, \breve{\U}]  =  
\frac{1}{2\pi {\rm i}} \, 
\oint \frac{{\rm d}\z}{\z} \,  
 \int 
 \rd^4 x\,{\rm d}^4\q
 \, 
K \big( \U^I (\z), \breve{\U}^{\bar{J}} (\z)  \big) ~.
\label{nact} 
\eea
The arctic $\U (\z)$ and  antarctic $\breve{\U} (\z)$ dynamical variables  
are generated by an infinite set of ordinary superfields:
\be
 \U (\z) = \sum_{n=0}^{\infty}  \, \U_n \z^n = 
\F + \S \,\z+ O(\z^2) ~,\qquad
\breve{\U} (\z) = \sum_{n=0}^{\infty}  \, {\bar
\U}_n
 (-\z)^{-n}~.
\label{exp}
\ee
Here $\F$ is chiral, $\S$  complex linear, 
eq. (\ref{chiral+linear}),
and the remaining component superfields are unconstrained complex 
superfields.  
The above  theory is a minimal $\cN=2$ extension of the
general four-dimensional $\cN=1$ supersymmetric 
nonlinear $\s$-model \cite{Zumino}
\be
S[\F, \bar \F] =  \int 
\rd^4 x\,{\rm d}^4\q
\, K(\Phi^{I},
 {\bar \Phi}{}^{\bar{J}})  ~,
\label{nact4}
\ee
with $K$ the  K\"ahler potential of a real analytic K\"ahler manifold $\cM$.

The study of $\s$-models of the form (\ref{nact})
was initiated in \cite{K98,GK1,GK2} because of their interesting geometric properties.
They form  a subset in the family of most general hypermultiplet theories
in projective superspace \cite{LR-projective} 
obtained by replacing  $K \big( \U , \breve{\U}   \big)$ in (\ref{nact})
with a Lagrangian $ K \big( \U , \breve{\U} , \z  \big)$ with explicit dependence on $\z$
(geometric aspects  of these  most general $\s$--models are briefly discussed in \cite{LR2008}). 
Our primary interest in such theories in the present  paper is motivated by  the fact 
that the off-shell $\cN=2$  superconformal $\s$-models of weight-one polar hypermultiplets
\cite{K-hyper} constitute a subclass in the family of actions (\ref{nact}). 

\subsection{General properties}
The $\cN=2$  supersymmetric  nonlinear $\s$-model  (\ref{nact}) 
inherits  all the geometric features of
its $\cN=1$ predecessor (\ref{nact4}). 
The K\"ahler invariance of the latter,
$K(\F, \bar \F) \to K(\F, \bar \F) +
\L(\F)+  {\bar \L} (\bar \F) $,
turns into 
\be
K(\U, \breve{\U})  \quad \longrightarrow \quad K(\U, \breve{\U}) ~+~
\L(\U) \,+\, {\bar \L} (\breve{\U} ) 
\label{kahl2}
\ee
for the model (\ref{nact}).
A holomorphic reparametrization of the K\"ahler manifold,
$ \F^I  \to \F'{}^I=f^I \big( \F \big) $,
has the following
counterpart
\be
\U^I (\z) \quad  \longrightarrow  \quad \U'{}^I(\z)=f^I \big (\U(\z) \big)
\label{kahl3}
\ee
in the $\cN=2$ case. Therefore, the physical
superfields of the 
${\cal N}=2$ theory
\be
 \U^I (\z)\Big|_{\z=0} ~=~ \F^I ~,\qquad  \quad \frac{ {\rm d} \U^I (\z) 
}{ {\rm d} \z} \Big|_{\z=0} ~=~ \S^I ~,
\label{kahl4} 
\ee
should be regarded, respectively, as  coordinates of a point in the K\" ahler
manifold and a tangent vector at  the same point. 
Thus the variables $(\F^I, \S^J)$ parametrize the holomorphic tangent 
bundle $T\cM$ of the K\"ahler manifold $\cM$ \cite{K98}. 

To describe the theory in terms of 
the physical superfields $\F$ and $\S$ only, 
all the auxiliary 
superfields have to be eliminated  with the aid of the 
corresponding algebraic equations of motion
\bea
\oint \frac{{\rm d} \z}{\z} \,\z^n \, \frac{\pa K(\U, \breve{\U} ) }{\pa \U^I} 
~ = ~ \oint \frac{{\rm d} \z}{\z} \,\z^{-n} \, 
\frac{\pa K(\U, \breve{\U} ) } {\pa \breve{\U}^{\bar J} } 
~ = ~ 0 ~, \qquad n \geq 2 ~ .               
\label{asfem}
\eea
Let $\U_*(\z) \equiv \U_*( \z; \F, {\bar \F}, \S, \bar \S )$ 
denote a unique solution subject to the initial conditions
\bea
\U_* (0)  = \F ~,\qquad  \quad \dt{\U}_* (0) 
 = \S ~.
\label{geo3} 
\eea
The auxiliary superfields $\U_2, \U_3, \dots$, and their
conjugates,  can be eliminated  in perturbation theory  
using the ansatz \cite{KL}
\bea
\U^I_n = 
\sum_{p=0}^{\infty} 
G^I{}_{J_1 \dots J_{n+p} \, \bar{L}_1 \dots  \bar{L}_p} (\F, {\bar \F})\,
\S^{J_1} \dots \S^{J_{n+p}} \,
{\bar \S}^{ {\bar L}_1 } \dots {\bar \S}^{ {\bar L}_p }~, 
\qquad n\geq 2~.
\label{ansatz}
\eea
Assuming that 
the auxiliary superfields 
have been eliminated, 
the action (\ref{nact}) should take the form
 \cite{GK1,GK2}:
\bea
S_{{\rm tb}}[\F,  \S]  &=&  
\frac{1}{2\pi {\rm i}} \, \oint \frac{{\rm d}\z}{\z} \,  
 \int \rd^4 x\,{\rm d}^4\q \, 
K \big( \U_* (\z), \breve{\U}_* (\z)  \big) \non \\
&=& \int 
\rd^4 x\,{\rm d}^4\q
\, \Big\{
K \big( \F, \bar{\F} \big)+  
\cL \big(\F, \bar \F, \S , \bar \S \big)\Big\}~,\non \\
\cL  \big(\F, \bar \F, \S , \bar \S \big)
&=&
\sum_{n=1}^{\infty}  \cL_{I_1 \cdots I_n {\bar J}_1 \cdots {\bar 
J}_n }  \big( \F, \bar{\F} \big) \S^{I_1} \dots \S^{I_n} 
{\bar \S}^{ {\bar J}_1 } \dots {\bar \S}^{ {\bar J}_n }
\equiv \sum_{n=1}^{\infty}  \cL^{(n)} 
~,~~~~~~~~
\label{act-tab}
\eea
where $\cL_{I {\bar J} }=  - g_{I \bar{J}} \big( \F, \bar{\F}  \big) $ 
and the coefficients $\cL_{I_1 \cdots I_n {\bar J}_1 \cdots {\bar 
J}_n }$, for  $n>1$, 
are tensor functions of the K\"ahler metric
$g_{I \bar{J}} \big( \F, \bar{\F}  \big) 
= \pa_I 
\pa_ {\bar J}K ( \F , \bar{\F} )$,  the Riemann curvature $R_{I {\bar 
J} K {\bar L}} \big( \F, \bar{\F} \big) $ and its covariant 
derivatives.  Each term in the action contains equal powers
of $\S$ and $\bar \S$, since the original model (\ref{nact}) 
is invariant under rigid U(1)  transformations
 \cite{GK1}
\be
\U(\zeta) ~~ \mapsto ~~ \U({\rm e}^{{\rm i} \a} \zeta) 
\quad \Longleftrightarrow \quad 
\U_n(z) ~~ \mapsto ~~ {\rm e}^{{\rm i} n \a} \U_n(z) ~.
\label{rfiber}
\ee
${}$For illustration, we give the explicit expressions \cite{K-hyper}
for two next-to-leading   terms appearing in the expansion of $\cL$:
\begin{subequations}
\bea
\cL^{(2)} &=& \frac{1}{4} R_{I_1 {\bar J}_1 I_2 {\bar J}_2} \S^{I_1}\S^{I_2}
{\bar \S}^{{\bar J}_1}{\bar \S}^{{\bar J}_2}~,
\label{L2}   \\
\cL^{(3)} &=&- \frac{1}{12} \Big\{ \frac{1}{6} 
\{ \nabla_{I_3}, { \nabla}_{{\bar J}_3} \}
R_{I_1 {\bar J}_1 I_2 {\bar J}_2} 
+R_{I_1 {\bar J}_1 I_2 }{}^LR_{L {\bar J}_2 I_3 {\bar J}_3}\Big\}   
\S^{I_1}\S^{I_2} \S^{I_3}
{\bar \S}^{{\bar J}_1} {\bar \S}^{{\bar J}_2} {\bar \S}^{{\bar J}_3}~.~~~~~~
\label{L3}
\eea
\end{subequations}
The expression for $\cL^{(4)}$, which is somewhat messy, can be found  given in \cite{K-hyper}.

\subsection{Considerations of extended supersymmetry} 
\label{2.2}
The action (\ref{nact}) is manifestly $\cN=1$ supersymmetric, and is also
invariant under the off-shell  second supersymmetry 
transformation \cite{LR-projective} (see also \cite{K-hyper} for a detailed derivation): 
\begin{subequations}
\bea
\d \U_0 &=& {\bar \ve}_{\dt \a} {\bar D}^{\dt \a} \U_1 ~,
\qquad 
\d \U_1 =-\ve^\a D_\a \U_0 
+   {\bar \ve}_{\dt \a}{\bar D}^{\dt \a} \U_2 
~,  \label{arctic7} \\
\d \U_k &=&-\ve^\a D_\a \U_{k-1} 
+{\bar \ve}_{\dt \a} {\bar D}^{\dt \a} \U_{k+1} 
~, \qquad k>1~.
\label{arctic8} 
\eea
\end{subequations}
Upon elimination of the auxiliary superfields, 
this symmetry turns into the following:
\bea
\d \F &=& {\bar \ve}_{\dt \a} {\bar D}^{\dt \a} \S ~,
\qquad 
\d \S =-\ve^\a D_\a \F 
+   {\bar \ve}_{\dt \a}{\bar D}^{\dt \a} \U_2 \big(\F, \bar \F, \S , \bar \S \big)
~,  
\label{arctic9} 
\eea
where $\U_2$ now a composite field of the general form given in (\ref{ansatz}).
Since $\U_2$ transforms as a connection under 
the holomorphic reparametrizations (\ref{kahl3}),
\bea
\U_2^I  \quad  \longrightarrow  \quad 
\U'{}^I_2 
= \hf  \frac{ \pa^2 f^I  \big( \F \big) }{\pa \F^J \pa \F^K}\, \S^J \S^K
+ \frac{ \pa f^I  \big( \F \big) }{\pa \F^J }\, \U_2^J ~,
\eea
we can rewrite $\U_2$ in more specific form \cite{K-hyper}:
\bea
\U^I_2 (\F,  \bar \F, \S , \bar \S)&=& -\hf \G^I_{JK} \big( \F, \bar{\F} \big) \, \S^J\S^K+
G^I(\F,  \bar \F, \S , \bar \S) ~, \non \\
G^I(\F,  \bar \F, \S , \bar \S)&:=&
\sum_{p=1}^{\infty} 
G^I{}_{J_1 \dots J_{p+2} \, \bar{L}_1 \dots  \bar{L}_p} (\F, {\bar \F})\,
\S^{J_1} \dots \S^{J_{p+2}} \,
{\bar \S}^{ {\bar L}_1 } \dots {\bar \S}^{ {\bar L}_p }~,~~~~~ 
\label{ansatz2}
\eea
with $\G^I_{JK} 
( \F , \bar{\F} )$  the Christoffel symbols for the  
K\"ahler metric $g_{I \bar J} ( \F , \bar{\F} )$. 
Here the coefficients $G^I{}_{J_1 \dots J_{p+2} \, \bar{L}_1 \dots  \bar{L}_p} (\F, {\bar \F})$
are tensor functions of the K\"ahler metric,
the Riemann curvature $R_{I {\bar 
J} K {\bar L}} \big( \F, \bar{\F} \big) $ and its covariant 
derivatives. To leading order, $G^I$ is  \cite{K-hyper}:
\bea
G^I(\F,  \bar \F, \S , \bar \S) = \frac{1}{6} \nabla_{J_1}R_{J_2 {\bar L} J_3 }{}^I (\F , \bar \F )\,
\S^{J_1}\S^{J_2} \S^{J_3}{\bar \S}^{{\bar L}} +\cO \big( \S^4  \bar \S^2 \big)~.
\eea

${}$For the action (\ref{act-tab}) to be invariant under the supersymmetry transformations
(\ref{arctic9}) and (\ref{ansatz2}), it can be shown that
there should exist a function 
\bea
\X(\F,  \bar \F, \S , \bar \S)&:=&
\sum_{n=2}^{\infty} 
\X_{I_1 \dots I_{n+1} \, \bar{J}_1 \dots  \bar{J}_n} (\F, {\bar \F})\,
\S^{I_1} \dots \S^{I_{n+1}} \,
{\bar \S}^{ {\bar J}_1 } \dots {\bar \S}^{ {\bar J}_n }~,
\label{Xi}
\eea
with tensor coefficients $\X_{I_1 \dots I_{n+1} \, \bar{J}_1 \dots  \bar{J}_n} (\F, {\bar \F})$,
such that the following equations hold:
\begin{subequations} 
\bea
\frac{\pa \cL}{\pa \S^J} \,\frac{\pa G^J}{\pa {\bar \S}^{\bar I} } 
&=& \frac{ \pa \X}{ \pa {\bar \S}^{\bar I} } ~, 
\label{master1}\\
\nabla_I \cL+ \frac{\pa \cL}{\pa \S^J} \,\frac{\pa G^J}{\pa { \S}^{I} } 
&=& \frac{ \pa \X}{ \pa { \S}^{ I} } ~,
\label{master2} \\
\hf R_{K {\bar I} L }{}^J\, \frac{\pa \cL}{\pa \S^J}\, \S^K \S^L
+ \frac{\pa \cL}{\pa {\bar \S}^{\bar I} }
+g_{J \bar{I}}\, \S^J 
- \frac{\pa \cL}{\pa \S^J} \,
\nabla_{\bar I} G^J &=&-\nabla_{\bar I} \X~.
\label{master3}
\eea
\end{subequations}
Here we have defined 
\bea
\nabla_I \cL &:=&\sum_{n=1}^{\infty} 
\Big( \nabla_I \cL_{J_1 \cdots J_n {\bar L}_1 \cdots {\bar 
L}_n }  \big( \F, \bar{\F} \big)\Big)  \S^{J_1} \dots \S^{J_n} 
{\bar \S}^{ {\bar L}_1 } \dots {\bar \S}^{ {\bar L}_n }\non \\
&=& \frac{\pa \cL}{\pa \F^I} - \frac{\pa \cL}{\pa \S^K}\, \G^K_{IJ} \,\S^J~,
\eea
and similarly for $\nabla_{\bar I} G^J$ and $\nabla_{\bar I} \X$.

The objects under consideration have several useful properties:
\begin{subequations}
\bea
{ \S}^{ J} \frac{\pa \cL}{\pa { \S}^{ J} }  &=&
{\bar \S}^{\bar J}
\frac{\pa \cL}{\pa {\bar \S}^{\bar J} }  ~, 
\label{semi-homo1} \\
{ \S}^{ J} 
\frac{\pa G^I}{\pa { \S}^{ J} } &=&
{\bar \S}^{\bar J}
 \frac{\pa G^I}{\pa {\bar \S}^{\bar J} }  +2G^I~, 
 \label{semi-homo2}\\
   { \S}^{ J}
\frac{\pa \X}{\pa { \S}^{ J} } &=&
{\bar \S}^{\bar J}
 \frac{\pa \X}{\pa {\bar \S}^{\bar J} }  +\X~.
\label{semi-homo3}
\eea
\end{subequations} 
These properties and the equations (\ref{master1}) and (\ref{master2}) 
allow us to obtain the following expression for  $\X$:
\bea
\X = \S^I  \nabla_I \cL + 2 G^I\, \frac{\pa \cL}{\pa \S^I} ~.
\label{X1}
\eea
We see that $\X$ is uniquely determined in terms of the Lagrangian $\cL$ 
and the vector field $G^I$ appearing in the second supersymmetry transformation
(\ref{arctic9}), (\ref{ansatz2}). 

It is of interest to discuss the special case when $\cM$ is a Hermitian symmetric space
and thus the  curvature tensor is covariantly constant,
\be
\nabla_L  R_{I_1 {\bar  J}_1 I_2 {\bar J}_2}
= { \nabla}_{\bar L} R_{I_1 {\bar  J}_1 I_2 {\bar J}_2} =0~.
\label{covar-const}
\ee
This case has been studied in detail in \cite{AKL2,KN}, and therefore 
we can compare the above results  with those obtained in \cite{AKL2,KN}.
First of all, since the curvature is covariantly constant, we have 
\bea
\nabla_{ I} \cL =0~, \qquad G^I=0 \quad \Longrightarrow \quad \X=0~,
\label{covar-const2}
\eea
as a consequence of (\ref{X1}). The fact that the conditions (\ref{covar-const}) 
imply $G^I=0$ was noticed in \cite{K-hyper} and can be explained as follows.
The  tensor fields $G^I{}_{J_1 \dots J_{p+2} \, \bar{L}_1 \dots  \bar{L}_p \bar K} (\F, {\bar \F})$
in (\ref{ansatz2}) have an odd number of indices.  On the other hand, the K\"ahler metric,
its inverse  and the Riemann tensor are the only algebraically independent tensors 
in the case (\ref{covar-const}).
It is therefore easy to understand that any tensor descendant 
of these geometric objects must carry an even number of indices, 
and thus $G^I=0$.
If  the relations (\ref{covar-const2}) hold,  eq.
(\ref{master3}) reduces to the equation 
\bea
\hf R_{K {\bar I} L }{}^J\, \frac{\pa \cL}{\pa \S^J}\, \S^K \S^L
+ \frac{\pa \cL}{\pa {\bar \S}^{\bar I} }
+g_{J \bar{I}}\, \S^J  &=&0~
\label{linearequation}
\eea
derived in \cite{AKL2}. This equation allows one to uniquely reconstruct 
$\cL$ in the case of covariantly constant curvature \cite{AKL2,KN}.
Similarly, in the general case, eqs. (\ref{master1}) -- (\ref{master3}) can be used to determine
the functional form of $\cL$ and $G^I$.

\subsection{Dual formulation}
To construct a dual formulation of the theory (\ref{act-tab}), 
we follow \cite{GK1,GK2} and consider the first-order action
\bea
S_{\rm first-order}=   \int \rd^4 x\,{\rm d}^4\q \, 
\Big\{\,
K \big( \F, \bar{\F} \big)+  \cL\big(\F, \bar \F, \S , \bar \S \big)
+\J_I \,\S^I + {\bar \J}_{\bar I} {\bar \S}^{\bar I} 
\Big\}~.
\label{f-o}
\eea
Here the tangent vector $\S^I$ is complex unconstrained, 
while the one-form $\Psi_I$ is chiral, 
${\bar D}_{\dt \a} \J_I =0$.
Eliminating
$\S$'s and their conjugates, 
by using their equations of motion
\bea
\frac{\pa  }{\pa \S^I}  \cL\big(\F, \bar \F, \S , \bar \S \big)+ \J_I =0~,
\eea 
leads to the dual action
\bea
S_{{\rm ctb}}[\F,  \J]  
&=&   \int \rd^4 x\,{\rm d}^4\q \, 
\Big\{\,
K \big( \F, \bar{\F} \big)+    
\cH \big(\F, \bar \F, \J , \bar \J \big)\Big\}~,
\label{act-ctb}
\eea
where 
\bea
\cH \big(\F, \bar \F, \J , \bar \J \big)&=& 
\sum_{n=1}^{\infty} \cH^{I_1 \cdots I_n {\bar J}_1 \cdots {\bar 
J}_n }  \big( \F, \bar{\F} \big) \J_{I_1} \dots \J_{I_n} 
{\bar \J}_{ {\bar J}_1 } \dots {\bar \J}_{ {\bar J}_n } ~,\non \\
\cH^{I {\bar J}} \big( \F, \bar{\F} \big) 
&=& g^{I {\bar J}} \big( \F, \bar{\F} \big) 
~.
\label{h}
\eea
The variables $(\F^I, \J_J)$ parametrize the cotangent 
bundle $T^* \cM$ of the K\"ahler manifold $\cM$ \cite{GK1}.
The superfield Lagrangian
\bea
{\mathbb K} (\F, \bar \F, \J, \bar \J )
:=K \big( \F, \bar{\F} \big)+  \cH(\F, \bar \F, \J, \bar \J )
\label{H-Kpotential}
\eea
is the hyperk\"ahler  potential of the target space.

It should be noted that eq.  (\ref{semi-homo1}) is equivalent to the relation
\bea
{\J}_{I} \frac{\pa \cH}{\pa {\J}_{ I} }  &=&
{\bar \J}_{\bar I}
\frac{\pa \cH}{\pa {\bar \J}_{\bar I} }  ~,
\label{semi-homo4} 
\eea
which will be used in what follows.
It follows from (\ref{semi-homo4}) 
that the group U(1) acts on  the hyperk\"ahler manifold $T^*\cM$ by holomorphic transformations
\bea
\F^I ~ \longrightarrow ~  \F^I~, 
\qquad 
\J_I ~ \longrightarrow ~  {\rm e}^{-{\rm i}  \a}\J_I~, \qquad \a \in {\mathbb R}~,
\eea
and this action is isometric with respect to the K\"ahler metric (see (\ref{nott}) for notation)
\bea
{\bm g}_{a \bar b} = \frac{\pa^2 \mathbb K}{  \pa \f^a   \pa {\bar \f}^{\bar b} }
= \left(\begin{array}{cc}
 \frac{\pa^2 \mathbb K}{\pa \F^I \pa {\bar \F}^{\bar J}} 
 ~ &   \frac{\pa^2 \mathbb K}{\pa \F^I \pa {\bar \J}_{\bar J}}  \\
 \frac{\pa^2 \mathbb K}{\pa \J_I \pa {\bar \F}^{\bar J}} 
~ &  \frac{\pa^2 \mathbb K}{\pa \J_I \pa {\bar \J}_{\bar J}} 
\end{array}
\right)~.
\eea
This agrees with results in the mathematical literature \cite{Kaledin,Feix}.

The first-order action (\ref{f-o}) can be shown to be invariant under 
the following second 
supersymmetry transformation:
\begin{subequations}
\bea
\d \F^I &=&\hf {\bar D}^2 \big\{ \overline{\e \q} \, \S^I\big\} ~,  \\
\d \S^I &=& -\e D \F^I -\hf   {\bar \e} {\bar D}
\Big\{ \G^I_{~JK} 
\, \S^J\S^K \Big\} 
-\hf  \overline{\e \q} \, \G^I_{~JK} 
 \, \S^J {\bar D}^2\S^K \non \\
&&+  {\bar \e} {\bar D} G^I  
+\hf 
\overline{\e \q} \, \frac{\pa G^I}{\pa \S^J}  \, \S^J {\bar D}^2\S^J 
~,  \\
\d \J_I &=&- \hf {\bar D}^2 \Big\{ \overline{\e \q} \, 
\Big(
K_I 
- \G^K_{~IJ} 
 \,\J_K \,
\S^J
+\frac{\pa G^J}{\pa \S^I}  \,\J_J +\frac{\pa \X}{\pa \S^I} \Big) \Big\}
~.
\label{SUSY3}
\eea
\end{subequations}
These results, in conjunction with eq. (\ref{master2}) and the standard properties
of Legendre transform, 
lead to the supersymmetry invariance of the dual theory (\ref{act-ctb}):
\begin{subequations}
\bea
\d \F^I &=&\phantom{-}\hf {\bar D}^2 \Big\{ \overline{\e \q} \, \frac{\pa \cH}{\pa \J_I} \Big\} ~, 
\label{SUSY-ctb1} \\
\d \J_I &=&- \hf {\bar D}^2 \Big\{ \overline{\e \q} \, 
\Big( K_I  
- \G^K_{~IJ} \,\J_K \, \frac{\pa \cH}{\pa \J_J}  
+\nabla_I \cH \Big) \Big\}~.
\label{SUSY-ctb2}
\eea
\end{subequations}
This form of the second supersymmetry  is useful in the special case when the K\"ahler 
manifold is Hermitian symmetric, eq. (\ref{covar-const}); then $\nabla_I \cH =0$, 
and the transformations (\ref{SUSY-ctb1}) and (\ref{SUSY-ctb2}) reduce to those obtained in \cite{AKL2}.
A different form for the second supersymmetry follows from the identity
\be
\nabla_I \cH =  \frac{\pa \cH}{\pa \F^I} 
+ \G^K_{~IJ} \,\J_K \, \frac{\pa \cH}{\pa \J_J}  
\ee 
and the explicit expression for the hyperk\"ahler potential, eq. (\ref{H-Kpotential}).
These observations lead to 
\bea
\d \F^I &=&\hf {\bar D}^2 \Big\{ \overline{\e \q} \, \frac{\pa \mathbb K}{\pa \J_I} \Big\} ~, 
\qquad
\d \J_I =- \hf {\bar D}^2 \Big\{ \overline{\e \q} \, 
\frac{\pa \mathbb K}{\pa \F^I}   \Big\}~.
\label{SUSY-ctb4}
\eea
Finally, if we introduce the condensed notation 
\be
\f^a := (\F^I\,, \J_I) ~, \qquad {\bar \f}^{\,\bar a} = ({\bar \F}^{\bar I}\,, {\bar \J}_{\bar I}), 
\label{nott}
\ee
as well as the standard symplectic matrix ${\mathbb J} =({\mathbb J}^{a b} )$, its inverse
${\mathbb J}^{-1} =(-{\mathbb J}_{a b} )$ and their complex conjugates,
\bea
{\mathbb J}^{a b} = {\mathbb J}^{\bar a \bar b} = 
\left(
\begin{array}{rc}
0 ~ &  {\mathbbm 1} \\
-{\mathbbm 1} ~ & 0  
\end{array}
\right)~,  \qquad 
{\mathbb J}_{a b} = {\mathbb J}_{\bar a \bar b} = 
\left( 
\begin{array}{rc}
0 ~ &  {\mathbbm 1} \\
-{\mathbbm 1} ~ & 0  
\end{array}
\right)~, 
\eea
then the supersymmetry transformation (\ref{SUSY-ctb4}) can be rewritten as 
\bea 
\d \f^a &=&\hf {\mathbb J}^{ab} \,
{\bar D}^2 \Big\{ \overline{\e \q} \, \frac{\pa \mathbb K}{\pa \f^b} \Big\} 
= \hf {\bar D}^2 \Big\{ \overline{\e \q} \, {\bar {\bm \O}}^a \Big\}
~, \qquad  {\bar {\bm \O}}^a := {\mathbb J}^{ab} \, \frac{\pa \mathbb K}{\pa \f^b}~.
\label{SUSY-ctb5}
\eea
The universal form of this transformation law  is quite remarkable.

\subsection{Holomorphic two-form}
By definition, the anti-holomorphic two-form is  
\bea
{\bar {\bm \o}}_{\bar b \bar c} = {\bm g}_{a \bar b} \,{\bar {\bm \O}}^a{}_{,\bar c}~,
\eea 
with $ {\bm g}_{a \bar b} $ the K\"ahler metric.
As is seen from (\ref{SUSY-ctb5}),  ${\bar {\bm \o}}_{\bar b \bar c} $ is indeed antisymmetric,
\bea
{\bar {\bm \o}}_{\bar a \bar b} 
={\bm g}_{ \bar a c}  \,{\mathbb J}^{cd} \,{\bm g}_{d \bar b}   ~.
 \eea

Direct calculations, based on the explicit structure of the hyperk\"ahker 
potential $\mathbb K$, show that the two-form ${\bm \o}_{ab}$ looks like 
\bea
{\bm \o}_{ab} = {\mathbb J}_{ab} + \cO (\J \bar \J )~. 
\non
\eea
Since ${\bm \o}_{ab}$ should be holomorphic, we  immediately conclude that
\bea
{\bm \o}_{ab} = {\mathbb J}_{ab}~, \qquad
{\bar {\bm \o}}_{\bar a \bar b} = {\mathbb J}_{\bar a \bar b}~.
\eea
We see that the holomorphic symplectic two-form
${\bm \o}^{(2,0)}$ of the hyperk\"ahler manifold $T^*\cM$ coincides with 
the canonical holomorphic symplectic two-form, 
\bea
{\bm \o}^{(2,0)} := \hf {\bm \o}_{ab} {\rm d}\f^a \wedge {\rm d} \f^b
 =  {\rm d} \F^I    \wedge{\rm d}\J_I~.
 \label{omega-j}
\eea
This agrees with results in the mathematical literature \cite{Kaledin,Feix}.

Next, since the metric is Hermitian with respect to each of the complex structures,
we conclude
\bea
{ {\bm \o}}^{ a  b} = {\bm g}^{a \bar  c}  {\bm g}^{b \bar  d} {\bar {\bm \o}}_{\bar c \bar d} 
= {\mathbb J}^{ab}~, \qquad 
{\bar {\bm \o}}^{\bar a \bar b} =  {\bm g}^{ \bar a c}  {\bm g}^{ \bar b d} {\bm \o}_{cd}
={\mathbb J}^{\bar a \bar b}~.
\eea
Since ${ {\bm \o}}^{ a  b} $ has been shown to coincide with the symplectic matrix ${\mathbb J}^{ab}$, 
the expression for ${\bar {\bm \O}}^a $
in our supersymmetry transformation law (\ref{SUSY-ctb5}) 
takes the form (\ref{barOmega}).

\section{Superconformal nonlinear sigma-models I}
\setcounter{equation}{0}

It was  demonstrated in \cite{K-hyper} that  the action (\ref{nact}) 
is $\cN=2$ superconformal  
provided:\\ 
(i) the arctic variables $\U^I (\z)$ transform
as superconformal weight-one arctic multiplets;  \\ 
(ii) the K\"ahler potential obeys the homogeneity condition
\bea
\F^I \frac{\pa}{\pa \F^I} K(\F, \bar \F) =  K( \F,   \bar \F)~.
\label{Kkahler2}
\eea
With the homogeneity condition imposed, no K\"ahler invariance survives. 
The geometric interpretation of such $\s$-models, 
albeit formulated in a slightly different form, was given in \cite{KLvU}.
This interpretation will be briefly discussed in section \ref{Discussion}.

\subsection{Off-shell superconformal transformations}

It was also shown in \cite{K-hyper} that the general $\cN=2$ superconformal 
transformation decomposes, upon reduction to $\cN=1$ superspace, 
into three types of $\cN=1$  transformations. The latter are the following:\\
${}\quad$ {\bf 1.} An arbitrary $\cN=1$ superconformal transformation generated by 
\bea
{\bm \x} = {\overline {\bm \x}} = {\bm \x}^a (z) \,\pa_a + {\bm \x}^\a (z)\,D_\a
+ {\bar {\bm \x}}_{\dt \a} (z)\, {\bar D}^{\dt \a}
\label{n=1scf1}
\eea  
such that 
\be
[{\bm \x} \;,\; D_\a ] 
= \bm { \o}_\a{}^\b  D_\b +
\Big({\bm \s} - 2 \bar{ {\bm \s}}  \Big) D_\a~, 
\label{n=1scf2}
\ee
see \cite{BK} for more details. 
This transformation acts on the components $\U_k$ of an weight-one arctic 
multiplet by the rule:
\bea
\d \U_k = -{\bm \x} \U_k - 2k(\bar{\bm \s} - {\bm \s})\U_k -2 {\bm \s}\U_k~.
\label{arctic3} 
\eea
${}\quad$ {\bf 2.} An extended  superconformal transformation
generated by  a spinor parameter $\r^\a$  constrained as
\bea
{\bar D}_{\dt \a} \r^\b =0~, \qquad D^{(\a}\r^{\b )}=0~,
\label{ro}
\eea
and hence
\be
\pa^{{\dt \a} (\a} \r^{\b )} = D^2 \r^\b =0~.
\ee
The general solution for (\ref{ro}) is 
 \bea
 \r^\a(x_{(+)}, \q) = \e^\a
+ \l\, \q^\a - {\rm i} \,{\bar \eta}_{\dt \a}\, x^{{\dt \a}\a}_{(+)}~, 
\qquad x^a_{(+)} = x^a +{\rm i} \q \s^a \bar \q~.
\label{ro-exp}
\eea
Here the {\it constant} parameters $\e^\a$, $\l$ and ${\bar \eta}_{\dt \a}$
correspond to  (i) a second Q-supersymmetry 
transformation $(\e^\a$); (ii) an off--diagonal SU(2)-transformation\footnote{In 
the standard $\cN=2$ superspace parametrized by variables
$z^A =(x^a, \q^\a_i, {\bar \q}_{\dt \a}^i)$,  this transformation rotates  the Grassmann variable 
$\q^\a_{\1}$ into $\q^\a_{\2}$ and vice versa. }
($\l$); 
and (iii) a second S-supersymmetry transformation (${\bar \eta}_{\dt \a}$). 
The extended superconformal transformation 
acts on the components $\U_k$ of an weight-one arctic 
multiplet by the rule:
\begin{subequations}
\bea
\d \U_0 &=& {\bar \r}_{\dt \a} {\bar D}^{\dt \a} \U_1 
+\hf \big( {\bar D}_{\dt \a} {\bar \r}^{\dt \a} \big) \U_1 ~,
\non \\
\d \U_1 &=&-\r^\a D_\a \U_0 
+  {\bar D}_{\dt \a}\big( {\bar \r}^{\dt \a}  \U_2 \big)
-\frac{1}{2}\big(D^\a\r_\a\big) \U_0 ~, 
\label{arctic5} \\
\d \U_k &=&-\r^\a D_\a \U_{k-1} 
+{\bar \r}_{\dt \a} {\bar D}^{\dt \a} \U_{k+1} \non \\
&&
+\hf (k-2)\big(D^\a\r_\a\big) \U_{k-1} 
+\hf (k+1) \big( {\bar D}_{\dt \a} {\bar \r}^{\dt \a} \big) \U_{k+1}
~, \qquad k>1~. 
\label{arctic6} 
\eea
\end{subequations}
${}\quad$ {\bf 3.} A shadow chiral rotation.
In $\cN=2$ superspace, this  is a phase transformation
of $\q^\a_{\2}$ only, with $\q^\a_{\1}$ kept unchanged.
Its action on the arctic weight-one multiplet is 
\bea
\U(\z) \quad &\longrightarrow & \quad \U'(\z)=
{\rm e}^{-({\rm i} /2) \a}\, \U({\rm e}^{{\rm i} \a} \z)~. 
\label{shadow4}
\eea

\subsection{Homogeneity conditions}

Suppose we have eliminated  all the auxiliary superfields $\U_2, \U_3, \dots$, and their
conjugates. Let $\U_*(\z) \equiv \U_*( \z; \F, {\bar \F}, \S, \bar \S )$ 
be the unique solution to the auxiliary field equations (\ref{asfem})
under  the initial conditions (\ref{geo3}).
Due to (\ref{Kkahler2} ), any rescaled holomorphic function 
\bea
\U_\clubsuit (\z):= c\, \U_* (\z) ~, \qquad c \in {\mathbb C} - \{ 0\} 
\eea
also solves the  auxiliary field equations (\ref{asfem})
and is characterized  by the   initial conditions  
\bea
\U_\clubsuit (0)  = c\, \F ~,\qquad  \quad \dt{\U}_\clubsuit (0) 
 = c\, \S ~.
\eea
This means that the Lagrangian $\cL \big(\F, \bar \F, \S , \bar \S \big)$ in  (\ref{act-tab})
obeys the homogeneity condition 
\bea
\Big( \F^I \frac{\pa}{\pa \F^I}  + \S^I \frac{\pa}{\pa \S^I} \Big) \cL \big(\F, \bar \F, \S , \bar \S \big)
= \cL \big(\F, \bar \F, \S , \bar \S \big)~.
\label{homo1}
\eea
Similar considerations imply that the functions $G^I \big(\F, \bar \F, \S , \bar \S \big)$ and 
$\X \big(\F, \bar \F, \S , \bar \S \big)$, which were introduced in subsection \ref{2.2},
also obey certain homogeneity conditions. In particular, one can notice  that 
\begin{subequations}
\bea
\Big(   {\bar \F}^{\bar J}   \frac{\pa }{\pa {\bar \F}^{\bar J} }
& +&    {\bar \S}^{\bar J}  \frac{\pa }{\pa {\bar \S}^{\bar J} } \Big) 
 G^I \big(\F, \bar \F, \S , \bar \S \big)= 0~,
\label{homo2} \\
 \Big(   {\bar \F}^{\bar J}  \frac{\pa }{\pa {\bar \F}^{\bar J} }
 &+&  {\bar \S}^{\bar J} \frac{\pa }{\pa {\bar \S}^{\bar J} }  \Big) 
 \X \big(\F, \bar \F, \S , \bar \S \big)= \X \big(\F, \bar \F, \S , \bar \S \big)~.
\label{homo3}
\eea
\end{subequations}
These relations can be used to obtain a new representation for $\X$. It is  
\bea
\X = - \frac{\pa \cL}{\pa {\bar \S}^{\bar I} } \,{\bar \F}^{\bar I} -K_I \,\S^I~,
\label{X-second}
\eea
compare with (\ref{X1}).

\subsection{On-shell superconformal transformations}

Upon elimination of the auxiliary superfields, 
the action (\ref{act-tab}) must be invariant under on-shell 
$\cN=2$ superconformal transformations of the physical superfields.
It is instructive to  check this invariance explicitly.

Let us consider the $\cN=1$ superconformal transformation 
(\ref{n=1scf1}),  (\ref{n=1scf2}).
The physical chiral $\F:= \U_0$ and complex linear 
$\S:=\U_1$ superfields transform as
\begin{subequations}
\bea
\d_{\bm \x}  \F &=& -{\bm \x} \F  -2{\bm \s} \F~, 
\label{arctic4-a} 
\\
\d_{\bm \x}  \S &=& -{\bm \x} \S - 2\bar{\bm \s} \S
~.
\label{arctic4} 
\eea
\end{subequations}
These transformations are consistent with the 
off-shell constraints ${\bar D}_{\dt \a} \F =0$ and ${\bar D}^2 \S =0$. 
Now, using the relations (\ref{semi-homo1}), (\ref{Kkahler2}) and (\ref{homo1}), 
one readily check that the action is invariant 
under the $\cN=1$ superconformal transformations,
\bea
\d_{\bm \x} S&=&   - \int \rd^4 x\,{\rm d}^4\q \, 
\Big( {\bm \x} +  2({\bm \s}+ \bar{\bm \s} ) \Big)
\Big\{\,
K \big( \F, \bar{\F} \big)+  
\cL \big(\F, \bar \F, \S , \bar \S \big)\Big\} \non \\
&=& - \int \rd^4 x\,{\rm d}^4\q \, (-1)^A 
\pa_A \Big( {\bm \x}^A \Big\{\,
K \big( \F, \bar{\F} \big)+  
\cL \big(\F, \bar \F, \S , \bar \S \big)\Big\} \Big)=0~.
\eea

We next turn to the extended superconformal transformation
\begin{subequations}
\bea
\d \F &=& {\bar \r}_{\dt \a} {\bar D}^{\dt \a} \S 
+\hf \big( {\bar D}_{\dt \a} {\bar \r}^{\dt \a} \big) \S ~,
\label{esct-onshell1} \\
\d \S &=&-\r^\a D_\a \F -\frac{1}{2}\big(D^\a\r_\a\big) \F
+  {\bar D}_{\dt \a}\Big( {\bar \r}^{\dt \a}  \U_2 \big(\F, \bar \F, \S , \bar \S \big)  \Big)~, 
\label{esct-onshell2}
\eea
\end{subequations}
with $\U_2 \big(\F, \bar \F, \S , \bar \S \big)$ given by eq. (\ref{ansatz2}).
It should be recalled that the parameter $\r^\a$ obeys the constraints (\ref{ro})
and its explicit form is given by (\ref{ro-exp}).
Since the parameters $\r^\a$ and ${\bar \r}^{\dt \a}$ are independent, modulo 
complex conjugation, it is sufficient to analyze the variation of the action 
in the case when $\r^\a =0$ and ${\bar \r}^{\dt \a} \neq 0$. It is 
\bea
\d_{\bar \r} S&=&    \int \rd^4 x\,{\rm d}^4\q \, 
 {\bar \r}_{\dt \a} {\bar D}^{\dt \a} \X \non \\
 &+&  \hf   \int \rd^4 x\,{\rm d}^4\q \, 
( {\bar D}_{\dt \a} {\bar \r}^{\dt \a} )\Big\{  \S^I \nabla_I\cL
+ 2 \frac{\pa \cL}{\pa { \S}^{ I} } \,G^I 
-  \frac{\pa \cL}{\pa {\bar \S}^{\bar I} } \,{\bar \F}^{\bar I} - K_I \,\S^I  \Big\}~.
\eea
Making use of the relations (\ref{X1}) and (\ref{X-second}) gives
\bea
\d_{\bar \r} S&=&    \int \rd^4 x\,{\rm d}^4\q \, 
{\bar D}_{\dt \a} \big(  {\bar \r}^{\dt \a} \X \big) =0~.
\eea

${}$Finally, it remains to consider the shadow chiral rotation (\ref{shadow4})
\bea
\F' = {\rm e}^{-({\rm i} /2) \a}\F~, \qquad 
\S' ={\rm e}^{({\rm i} /2) \a} \S~. 
\label{shadow5}
\eea
The action is  invariant under such transformations, 
as a consequence of  eq. (\ref{Kkahler2}), (\ref{semi-homo1}) and (\ref{homo1}).

\subsection{Dual formulation}
We now turn to considering the superconformal symmetries within
the dual formulation (\ref{act-ctb}). First of all, we should discuss the homogeneity 
properties of $ \cH \big(\F, \bar \F, \J , \bar \J \big)$.
Using the homogeneity condition (\ref{homo1})
and the standard properties of the Legendre transformation, 
one obtains 
\bea
\Big( \F^I \frac{\pa}{\pa \F^I}  +  {\bar \J}_{\bar I}   \frac{\pa }{\pa {\bar \J}_{\bar I} }
 \Big) \cH \big(\F, \bar \F, \J , \bar \J \big)
= \cH \big(\F, \bar \F, \J , \bar \J \big)~.
\label{homo4}
\eea
Taking into account eq. (\ref{semi-homo4}), this is equivalent to 
\bea
\Big( \F^I \frac{\pa}{\pa \F^I}  +  { \J}_{ I}   \frac{\pa }{\pa { \J}_{ I} }
 \Big) \cH \big(\F, \bar \F, \J , \bar \J \big)
= \cH \big(\F, \bar \F, \J , \bar \J \big)~.
\label{homo5}
\eea
 
The explicit structure of the first-order action, eq. (\ref{f-o}), as well as  the 
$\cN=1$ superconformal transformation law of $\S$, eq.  (\ref{arctic4}), imply that 
the corresponding transformation of $\J$ is
\bea
\d_{\bm \x}  \J &=& -{\bm \x} \J  -2{\bm \s} \J~
\label{N=1sc-Psi}
\eea 
which coincides with that of $\F$,  eq.  (\ref{arctic4}).
Using the homogeneity conditions (\ref{Kkahler2}) and (\ref{homo5}), 
one immediately sees that the action (\ref{act-ctb}) is $\cN=1$ superconformal.

As a next step, we should consider the extended superconformal transformation.
Using the transformation laws (\ref{esct-onshell1}) and (\ref{esct-onshell2}), 
one can check that    the first-order action (\ref{f-o})
possesses the following invariance:
\begin{subequations}
\bea
\d \F^I &=&\hf {\bar D}^2 \big\{ \bar {\bm \r} \, \S^I\big\} ~,  
\label{esc-dual1}\\
\d \S^I &=&  -\r D \F -\frac{1}{2}\big(D\r \big) \F
+  {\bar D}_{\dt \a}  \Big\{ {\bar \r}^{\dt \a} 
\Big( G^I - \hf \G^I_{~JK} \, \S^J\S^K  \Big)\Big\} \non \\
&&
-\hf  \bar {\bm \r} \, \G^I_{~JK} 
 \, \S^J {\bar D}^2\S^K 
+\hf \bar {\bm \r} \, \frac{\pa G^I}{\pa \S^J}  \, \S^J {\bar D}^2\S^J 
~,  
\label{esc-dual2}\\
\d \J_I &=&- \hf {\bar D}^2 \Big\{ \bar {\bm \r} \, 
\Big(K_I - \G^K_{~IJ}  \,\J_K \,\S^J
+\frac{\pa G^J}{\pa \S^I}  \,\J_J +\frac{\pa \X}{\pa \S^I} \Big) \Big\}
~.
\label{esc-dual3}
\eea
\end{subequations}
Here we have introduced the parameter $\bar{ \bm \r} $  and its conjugate defined by
\be
{\bar \r}_{\dt \a} = {\bar D}_{\dt \a} \bar {\bm \r}~, \qquad 
{ \r}_{\a} = { D}_{ \a}  {\bm \r}~.
\ee
Modulo redefinitions of the form
\be
 \bar {\bm \r}\quad \longrightarrow \quad  \bar {\bm \r} + \m ~, \qquad 
 {\bar D}_{\dt \a}\m =0~,
 \ee
the  parameter $ {\bm \r}$  can be chosen to be
\be
\bm \r = \t  + \e \q + \hf \l \, \q^2 +{\rm i} \, x^a \q \s_a {\bar \eta}~,
\label{rho-scalar}
\ee
with $\t$ corresponding to a ``central charge'' transformation.

It follows from eqs. (\ref{esc-dual1}) -- (\ref{esc-dual3}) that 
the dual theory (\ref{act-ctb}) is invariant under the following
extended superconformal transformation:
\begin{subequations}
\bea
\d \F^I &=&\phantom{-}\hf {\bar D}^2 \Big\{ \bar {\bm \r} \, \frac{\pa \cH}{\pa \J_I} \Big\} ~, 
\label{esc-dual4} \\
\d \J_I &=&- \hf {\bar D}^2 \Big\{ \bar {\bm \r} \, 
\Big( K_I  
- \G^K_{~IJ} \,\J_K \, \frac{\pa \cH}{\pa \J_J}  
+\nabla_I \cH \Big) \Big\}
~.
\label{esc-dual5}
\eea
\end{subequations}
Similarly to the second supersymmetry transformation (\ref{SUSY-ctb5}), 
this can be rewritten in terms of the hyperk\"ahler potential as follows:
\bea 
\d \f^a &=&\hf {\mathbb J}^{ab} \,
{\bar D}^2 \Big\{ \bar{\bm \r} \, \frac{\pa \mathbb K}{\pa \f^b} \Big\} 
~.
\label{esc-ctb}
\eea
If only the second term in the parameter (\ref{rho-scalar}) is kept, 
${ \bm \r} \to  \e \q $, then our transformation law (\ref{esc-ctb}) reduces to
(\ref{SUSY-ctb5}).

It remains to point out that the shadow chiral rotation (\ref{shadow5}) turns into 
\bea
\f^a ~ \longrightarrow ~  {\rm e}^{-({\rm i} /2) \a}\f^a~. 
\label{shadow6}
\eea

\subsection{Homothetic conformal Killing vector}
Before we continue, it is worth recalling the salient facts about  
homothetic conformal Killing vectors (see \cite{GR,deWRV} for more details).
By definition, a homothetic conformal Killing vector $\c$
on a K\"ahler manifold $(\cM, g_{a\bar b})$, 
\bea
\c = \c^a \frac{\pa}{\pa \f^a} + {\bar \c}^{\bar a}  \frac{\pa}{\pa {\bar \f}^{\bar a}}
\equiv \c^\m \frac{\pa}{\pa \vf^\m} ~,
\eea
obeys the constraint
\bea
\nabla_\n \c^\m = \d_\n{}^\m \quad \Longleftrightarrow \quad 
\nabla_b \c^a = \d_b{}^a~, \qquad 
\nabla_{\bar b} \c^a = \pa_{\bar b} \c^a = 0~.
\label{hcKv}
\eea
In particular,  $\c $ is holomorphic. Its properties include:
\bea
{ g}_{a \bar b} \, \c^a {\bar \c}^{\bar b} ={ K}~, 
\qquad \c_a := {g}_{a \bar b} \,{\bar \c}^{\bar b} = \pa_a { K}~,
\label{hcKv-pot}
\eea
with $K$ the K\"ahler potential.

If $ I^\m{}_\n$ is the complex structure on the K\"ahler manifold,  
then  $\u^\m  := I^\m{}_\n \, \c^\n$ is a holomorphic Killing vector.
In the case that  $\cM$ is  a hyperk\"ahler cone, there are three complex structure, 
$({J}_A){}^\m{}_\n $, and the Killing vectors $\u_A{}^\m:= ({J}_A){}^\m{}_\n \c^\n$ 
generate the Lie algebra of the  group SU(2).

Let us return to the hyperk\"ahler cone studied in the main body of this section.
Here the homothetic conformal Killing vector  vector proves to be 
\be
\c^a = \f^a~, \qquad {\bar \c}^{\bar a}={\bar \f}^{\bar a}~.
\label{hcKv-ctb}
\ee
This can readily be checked using the homogeneity condition (\ref{homo5}) 
and the explicit form of the Christoffel symbols on K\"ahler manifolds.

The extended superconformal transformation (\ref{esc-ctb}) can now be rewritten 
in terms of the homothetic conformal Killing vector:
\bea 
\d \f^a &=&\hf
{\bar D}^2 \Big\{ \bar{\bm \r} \,  {\bm \o}^{ab} \c_b \Big\} 
~.
\label{esc-ctb2}
\eea

\section{Superconformal nonlinear sigma-models II}
\setcounter{equation}{0}
We now have all prerequisites available 
to develop a chiral formulation in $\cN=1$ superspace for the most general 
$\cN=2$ superconformal nonlinear $\s$-model. Given a hyperk\"ahler cone $\cM$, 
we pick one of its complex structures, say $J_3$, and introduce complex 
coordinates $\f^a$ compatible with it. In these coordinates, $J_3$ has the form 
(\ref{J3}).  Two other complex structures, $J_1$ and $J_2$, become
\bea
J_1 = \left(
\begin{array}{cc}
0  ~ & ~ {g}^{a \bar c} {\bar \o}_{\bar c \bar b} \\
{g}^{ \bar a c } { \o}_{ c  b}
 ~ &   0
\end{array}
\right)~, \qquad 
J_2 = \left(
\begin{array}{cc}
0  ~ &  {\rm i}\,   {g}^{a \bar c} {\bar \o}_{\bar c \bar b}  \\
-{\rm i}\,   {g}^{ \bar a c } { \o}_{ c  b}~ &   0
\end{array}
\right)~,
\label{J1J2-hkc}
\eea
where ${g}_{a \bar b} $ be the K\"ahler metric, and 
$\o_{ab}$ the holomorphic symplectic two-form. 
Let  $\c$ be the homothetic conformal Killing vector, eq. (\ref{hcKv}).
We then have ${g}_{a \bar b} =\pa_a \pa_{\bar b} {K}$, where the potential $K$
is related to $\c$ according to  eq. (\ref{hcKv-pot}). 
With this K\"ahler potential, the $\s$-model action
(\ref{N=1sigma-model}) turns out to be $\cN=2$ superconformal, 
as we are going to prove. 

Our first observation is that  the action (\ref{N=1sigma-model}) is invariant under $\cN=1$
superconformal transformations of the form:
\bea
\d_{\bm \x}  \f^a &=& -{\bm \x} \f^a  -2{\bm \s}\, \c^a (\f)~, 
\eea
compare with  eqs. (\ref{arctic4-a}) and (\ref{N=1sc-Psi}). The invariance follows from the identity 
\be
\c^a (\f) \,\pa_a K (\f, \bar \f)  = K (\f , \bar \f) 
\label{idei}
\ee
and the standard properties of the $\cN=1$ superconformal Killing vectors \cite{BK}.
What we have actually demonstrated here, can be rephrased as follows. If a K\"ahler manifold $\cM$
possesses a homothetic conformal Killing vector $\c$, then 
the associated $\cN=1$ nonlinear $\s$-model  (\ref{N=1sigma-model}) is superconformal
(see also \cite{GR,KW}).

Next, we define the extended superconformal transformation  of $\f^a$:
\bea 
\d \f^a &=&\hf
{\bar D}^2 \Big\{ \bar{\bm \r} \,  { \o}^{ab} \c_b \Big\} ~,
\label{hkc-esc}
\eea
which should be compared with (\ref{esc-ctb2}). We now prove that the action is 
invariant under (\ref{hkc-esc}). As before, it is sufficient to evaluate the variation 
$\d_{\bar {\bm \r} }S$ which corresponds to the choice $\bm \r =0$ and $\bar {\bm \r} \neq 0$.
The variation of the action is 
\bea
\d_{\bar {\bm \r}} S&=& -\hf    \int \rd^4 x\,{\rm d}^4\q \, 
\big({\bar D}_{\dt \a} \c_a \big)
\big(  {\bar D}^{\dt \a} {\bm \r} \big) \, \o^{ab}\c_b
= -\hf    \int \rd^4 x\,{\rm d}^4\q \, {\bar \r}_{\dt \a} \big({\bar D}^{\dt \a} {\bar \f}^{\bar c} \big)
g_{\bar c\, a}  \, \o^{ab}\c_b \non \\
&=&  -\hf    \int \rd^4 x\,{\rm d}^2\q {\rm d}^2{\bar \q}
\, {\bar \r}_{\dt \a} \big({\bar D}^{\dt \a} {\bar \f}^{\bar a} \big)
{\bar \o}_{\bar a \bar b} \,{\bar \c}^{\bar b}~.
\eea
Since the tensor fields ${\bar \o}_{\bar a \bar b} $ and ${\bar \c}^{\bar b}$ are anti-holomorphic, 
and the parameter $ {\bar \r}_{\dt \a} $ is antichiral, 
the combination $ {\bar \r}_{\dt \a} {\bar \o}_{\bar a \bar b} \,{\bar \c}^{\bar b}$
appearing in the integrand is antichiral. 
As a result, doing the Grassmann integral $\int {\rm d}^2\q$ gives
\bea
\d_{\bar {\bm \r}} S&=& \frac{1}{8}  \int \rd^4 x\,{\rm d}^2{\bar \q}\,
 {\bar \r}_{\dt \a} {\bar \o}_{\bar a \bar b} \,{\bar \c}^{\bar b}\,D^2 {\bar D}^{\dt \a} {\bar \f}^{\bar a} 
=0~, 
\eea
for $D^2 {\bar D}^{\dt \a} {\bar \F} $ is identically zero for any antichiral superfield $\bar \F$.

${}$Finally, we define the infinitesimal shadow chiral rotation of $\f^a$: 
\bea
\d \f^a = -\frac{\rm i}{2} \a \, \c^a (\f) ~, \qquad \bar \a = \a~,
\eea
compare with (\ref{shadow6}). Because of the identity (\ref{idei}), this 
transformation leaves  the action invariant.
It should be remarked that the shadow chiral rotation is generated 
by the Killing vector 
\bea
\u = {\rm i} \,\c^a (\f) \frac{\pa}{\pa \f^a} -{\rm i} \, {\bar \c}^{\bar a} (\bar \f)  \frac{\pa}{\pa {\bar \f}^{\bar a}} ~.
\eea

\section{Discussion and future directions}
\setcounter{equation}{0}
\label{Discussion}

In this paper we  developed the formulation in terms of $\cN=1$ chiral superfields
for the following $\cN=2$ supersymmetric theories: (i) $\s$-models on cotangent bundles 
of  K\"ahler manifolds; (ii) general superconformal $\s$-models described by weight-one 
polar supermultiplets. Using the considerations of duality, we demonstrated that 
the holomorphic symplectic two-form ${\bm \o}_{ab}$  of 
the hyperk\"ahler space  $T^*\cM$  coincides with the canonical
holomorphic symplectic structure, eq. (\ref{omega-j})
In the case (ii), we also determined the homothetic conformal Killing vector $\c$, eq. (\ref{hcKv-ctb}).
The explicit expressions for ${\bm\o}^{(2,0)}$ and $\c$ are necessary in the context of the $\cN=2$ 
superconformal quotient construction \cite{deWRV} allowing one to reduce the hyperk\"ahler cone
to the corresponding quaternion K\"ahler space.

We also presented, in section 5, the most general $\cN=2$ superconformal nonlinear $\s$-model
formulated in terms of $\cN=1 $ chiral superfields.
It would be interesting to compare the corresponding superconformal transformations, 
obtained using off-shell techniques, with those defined in 
the component approach to $\cN=2$ superconformal $\s$-models \cite{deWKV}.

The off-shell $\cN=2$ superconformal $\s$-model defined by eqs. (\ref{nact}) and  (\ref{Kkahler2})
possesses a  slightly different realization \cite{K-hyper}. It can be defined if
the arctic weight-one hypermultiplets $\U^I (\z)$ 
include a compensator $\U(\z)$, that is an arctic multiplet with its 
lowest-order ($\z$-independent) 
component $\F$ is everywhere non-vanishing. 
Then, we can introduce new dynamical variables comprising the only
{\it weight-one} multiplet $\U(z,\z)$ and some set of {\it weight-zero} arctic multiplets
$\u^i(z,\z)$.
\bea
S [ \U (\z) , \u (\z)]&=&  
   \oint \frac{{\rm d}\z}{2\pi {\rm i} \z} \,  
 \int  {\rm d}^4 x \, {\rm d}^4\q\, \breve{\U} \,\U \,
 \exp \Big\{  \cK ( \u^i , \breve{\u}{}^{\bar j}   ) \Big\} ~,
\label{act-dual} 
\eea
with $ \cK(\u, \breve{\u} )$ the K\"ahler potential
of a K\"ahler manifod $\cM_K$.    
This action is invariant under K\"ahler transformations of the form
\bea
\U ~\longrightarrow ~{\rm e}^{-\L(\u)} \, \U~, \qquad 
\cK(\u, \breve{\u} ) ~\to ~ 
\cK(\u, \breve{\u} )+ \L(\u) + {\bar \L}( \breve{\u})~,
\eea
with $\L $ a holomorphic function. In accordance with 
\cite{KLvU} (see also \cite{HitchinKLR}),   
the space $\cM_K$ is necessarily  a  K\"ahler-Hodge manifold, 
and the  arctic variables $\U^I$ and $\X$ in (\ref{act-dual})  parametrize 
a holomorphic line bundle over  $\cM_K$.

As discussed in detail in \cite{KLvU}, the theory (\ref{act-dual}) possesses a dual formulation
in which the arctic compensator $\U$ and its conjugate are dualized 
into an $\cO(2)$ multiplet  $H(\z)$  \cite{KLR} (or $\cN=2$ tensor multiplet \cite{Wess}) 
\bea
H(\z) = \frac{1}{\z}\, \vf + G - \z \,{\bar \vf}~, \qquad {\bar D}_{\dt \a} \vf =0~, 
\qquad {\bar D}^2 G =0~, \quad {\bar G}=G~.
\label{tensor-series}
\eea
The corresponding action, which is a rigid supersymmetric version of the theory introduced in 
\cite{K-dual},  is
\bea 
S [ H(\z) , \u (\z)]= - \oint \frac{{\rm d}\z}{2\pi {\rm i} \z} \,  
 \int  {\rm d}^4 x \,{\rm d}^4\q\,
 H \ln H 
 + \oint \frac{{\rm d}\z}{2\pi {\rm i} \z} \,  
\int  {\rm d}^4 x \, {\rm d}^4\q\,H \,  \cK \big( \u^i, \breve{\u}{}^{\bar j}   \big) ~.
\label{act-dual-dual} 
\eea
Here the first term is the $\cN=2$ projective-superspace formulation \cite{KLR}
of the $\cN=2$ improved  tensor multiplet model \cite{deWPV}.

It would very interesting to extend the analysis given in sections 3 and 4 to the 
case of the $\s$-models  (\ref{act-dual}) and (\ref{act-dual-dual}).
The important feature of (\ref{act-dual}) is that the K\"ahler potential
$ \cK$ is essentially arbitrary. 
The structure of the action (\ref{act-dual}) is similar to that describing locally supersymmetric 
$\s$-modesl in $\cN=1$ supergravity, see e.g. \cite{BK} for a review.
As to the formulation (\ref{act-dual-dual}), it  turns out to be  useful for 
generating new hyperk\"ahler cones  \cite{KLvU}.

Our derivation of the chiral formulation in $\cN=1$ superspace for 
general $\cN=2$ superconformal nonlinear $\s$-models can naturally be extended 
to five and six dimensions. In the 5D  case, one has to use the formalism of superconformal 
Killing vectors and the off-shell superconformal $\s$-models introduced in \cite{K-hyper}.
In the 6D case, the general aspects of superconformal symmetry in superspace have been elaborated
by Park \cite{Park}. General off-shell 6D $\cN=(1,0)$ superconformal nonlinear $\s$-models 
have not yet been described in the literature, however they  can be constructed 
in complete analogy with the 5D construction of \cite{K-hyper}. 
It should be emphasized that 5D and 6D rigid supersymmetric nonlinear $\s$-models 
with eight supercharges have been formulated in 4D $\cN=1$ superspace in 
Refs. \cite{BX} and \cite{GPT-M}, respectively. The work of Bagger and Xiong 
was based on the careful analysis of supersymmetry transformations in five dimensions. 
As to the six-dimensional construction of \cite{GPT-M}, the hyperk\"ahler conditions 
on the target space geometry were derived by the authors on the basis of the 
requirement that  the  component action must be Lorentz invariant, without
any analysis of supersymmetry. 

In conclusion, we would like to raise an issue that is simplest to  formulate in terms of 
the $\cN=2$ supersymmetric $\s$-models on cotangent bundles of K\"ahler manfolds.
For these theories, we have considered the three different realizations:
(i) the off-shell $\U$-formulation  given by the action (\ref{nact});
(ii) the $\F \S$-formulation (\ref{act-tab}) which emerges  from (\ref{nact})
upon elimination of the auxiliary superfields; (iii) the $\F \J$-formulation (\ref{act-ctb})
which is dual to (\ref{act-tab}). Deriving the latter formulation was the actual goal of our 
analysis. So, do we really need the off-shell realization  (\ref{nact}) and/or its
reduced version (\ref{act-tab})? The answer is ``Yes'' in general. The point is that 
the ``Hamiltonian'' $\cH(\F, \bar \F, \J, \bar \J )$ in the hyperk\"ahler potential
(\ref{H-Kpotential}) must obey a highly nonlinear differential equation
(this formulation is therefore not suitable to study deformations of the hyperk\"ahler structure).
On the other hand, the tangent-bundle realization (\ref{act-tab}) is generated 
by two functions $\cL(\F, \bar \F, \S, \bar \S)$ and $G^I (\F, \bar \F, \S, \bar \S)$ 
which must satisfy the quadratic differential equations (\ref{master1}) -- (\ref{master3}), 
with $\X$ given by eq. (\ref{X1}). The problem of solving these equations is technically much simpler 
than that of solving the nonlinear equation obeyed by $\cH$.\footnote{Of course, 
the off-shell $\s$-model (\ref{nact}) is generated by an arbitrary real analytic function $K(\F, \bar \F)$, 
and therefore is most suitable if one is interested in deformations of the hyperk\"ahler structure.
But here we still have to address the problem of eliminating the auxiliary superfields.}
As an illustration, consider the case when $K(\F, \bar \F)$ corresponds to a Hermitian symmetric 
space. In this case, $G^I =\X=0$ and $\cL$ obeys the linear differential equation
(\ref{linearequation}) which is easy to solve \cite{AKL2,KN}. On the other hand, 
in the cotangent-bundle realization $\cH$ obeys the quadratic differential equation \cite{AKL2}
\be
\cH^I \,  g_{I {\bar J}} - \hf \, \cH^K\cH^L \,  R_{K {\bar J} L}{}^I \,\J_I =
{\bar \J}_{ \bar J} ~, 
\qquad \cH^I  = \frac{\pa \cH}{\pa \J_I} ~.
\label{cot-eq}
\ee
Solving this equation is more challenging. Its solution was found in \cite{KN}.
\\

\noindent
{\bf Acknowledgements:}\\
The hospitality of the Center for Quantum Spacetime at Sogang University
(grant number R11-2005-021) during the final stage of this project
is gratefully acknowledged. The author is also grateful to Ian McArthur for reading
the manuscript. This work is supported  in part by the Australian Research Council.

\appendix

\section{Extended supersymmetry} 
\setcounter{equation}{0}
Here we derive the conditions \cite{HKLR} for the $\s$-model (\ref{N=1sigma-model}) 
to be invariant under the transformations (\ref{LR-ansatz}) and (\ref{parameter}).
For our consideration, it is handy to make use of the following condensed notation:
\bea
S&=& \int {\rm d}^4 x {\rm d}^2 \q {\rm d}^2{\bar \q} \, K\big(\f^a, {\bar \f}^{\overline{b}}\big)
\equiv \int K~. 
\eea
Since the superfield parameters $\e$ and $\bar \e$ in (\ref{LR-ansatz}) are independent, 
modulo complex conjugation, it is sufficient  to analyze only the variation $\d_{\bar \e} S$
corresponding to the choice $\e =0$ and $\bar \e \neq 0$.

Varying the action gives
\bea 
\d_{\bar \e}S &=& \hf \int K_a {\bar D}^2  \big( {\bar \e} \,{\bar \O}^a \big)
= - \hf \int K_{a \overline{b} }\, \big( {\bar D}_{\dt \a} {\bar \f}^{\overline{b}} \big) \,
{\bar D}^{\dt \a}  \big( {\bar \e} \,{\bar \O}^a \big) \non \\
&=& - \hf \int g_{a \overline{b} }\, \big( {\bar D}_{\dt \a} {\bar \f}^{\overline{b}} \big) \,
\big\{  {\bar \O}^a  {\bar D}^{\dt \a}   {\bar \e} 
+ {\bar \e}\,  {\bar \O}^a{}_{,\overline{c} } \, {\bar D}^{\dt \a}  {\bar \f}^{\overline c} \, \big\} ~.
\label{vari1}
\eea
Choose $\bar \e = \bar \t = {\rm const}.$ 
Then $\d_{\bar \e}S$ reduces to 
\bea 
2 \d_{\bar \t}S &=& -   \bar \t \int g_{a \overline{b} }\, {\bar \O}^a{}_{,\overline{c} }
\big( {\bar D}_{\dt \a} {\bar \f}^{\overline{b}} \big) {\bar D}^{\dt \a}  {\bar \f}^{\overline c}
\equiv - \bar  \t \int {\bar \o}_{ \overline{b}  \overline{c} }
\big( {\bar D} {\bar \f}^{\overline{b}} \big) {\bar D}  {\bar \f}^{\overline c} 
~. 
\eea
Since $\big( {\bar D} {\bar \f}^{\overline{b}} \big) {\bar D}  {\bar \f}^{\overline c}$ is symmetric, 
the above variation vanishes if the equation (\ref{se1}) holds.
As a result, the variation (\ref{vari1}) becomes
\bea 
\d_{\bar \e}S &=&
 - \hf \int g_{a \overline{b} }\, \big( {\bar D}_{\dt \a} {\bar \f}^{\overline{b}} \big) \,
 {\bar \O}^a  {\bar D}^{\dt \a}   {\bar \e} ~.
\label{vari2}
\eea
One can now see that it is not necessary to constrain the parameter as in eq. 
(\ref{parameter}), and instead it is sufficient to impose the weaker constraint (\ref{parameter2}).
We see that  $\d_{\bar \e}S$  vanishes if and only if the following functional 
\bea
\X_{\dt \a} [\f, \bar \f] := \int  {\bar \O}^a 
g_{a \overline{b} }\, \big( {\bar D}_{\dt \a} {\bar \f}^{\overline{b}} \big) 
= \int  {\bar \O}^a \, {\bar D}_{\dt \a} K_{a } 
\label{functional}
\eea
is  identically zero. 

Let us vary (\ref{functional}) with respect to $\bar \f$: 
\bea
\d_{\bar \f}  
\X_{\dt \a} [\f, \bar \f]  &=& 
-2 \int {\bar \o}_{ \overline{b}  \overline{c} }\,
\d {\bar \f}^{\overline{b}}  {\bar D}_{\dt \a}  {\bar \f}^{\overline c} ~.
\eea
To make this variation vanish identically, one has to impose the equation (\ref{se2}), 
that is  ${\bar \o}_{ \overline{b}  \overline{c} } = {\bar \o}_{ \overline{b}  \overline{c} }(\bar \f )$.
Indeed, since $\bar \f$ and $\d \bar \f$ are  antichiral, we then have
\bea
 \int {\bar \o}_{ \overline{b}  \overline{c} }\,
\d {\bar \f}^{\overline{b}}  {\bar D}_{\dt \a}  {\bar \f}^{\overline c} 
=-\frac{1}{4}  \int {\rm d}^4 x {\rm d}^2{\bar \q} \, {\bar \o}_{ \overline{b}  \overline{c} }\,
\d {\bar \f}^{\overline{b}} 
D^2  {\bar D}_{\dt \a}  {\bar \f}^{\overline c}  =0~.
\non
\eea

Return to the general analysis of the requirement  that the functional (\ref{functional})
should vanish identically.
We can represent 
\bea
\X_{\dt \a} [\f, \bar \f] 
 = \frac{1}{16} \int {\rm d}^4 x  \, D^2 {\bar D}^2 \Big\{   {\bar \O}^a \, {\bar D}_{\dt \a} K_{a } \Big\} ~.
\label{D2barD2}
 \eea
Here ${\bar D}^2 \big\{   {\bar \O}^a \, {\bar D}_\ad K_{a } \big\} $ can be expressed in terms 
of the anti-holomorphic two-form ${\bar \o}_{ \overline{b}  \overline{c} }$, 
with the property that 
\be
D_\a {\bar \o}_{ \overline{b}  \overline{c} } =0~.
\ee
One thus obtains 
\bea 
{\bar D}^2 \Big\{   {\bar \O}^a \, {\bar D}_{\dt \a} K_{a } \Big\} 
&=& 2 {\bar \o}_{ \bar b  \bar c }  ({\bar D}_{\dt \a}  {\bar \f}^{\bar b}) {\bar D}^2
{\bar \f}^{\bar c}
+\hf  \cF_{\bar b ; \bar c \bar d}\,
({\bar D}_{\dt \a}  {\bar \f}^{\bar b}) ({\bar D}_{\dt \b}  {\bar \f}^{\bar c}) {\bar D}^{\dt \b}  
{\bar \f}^{\bar d} ~,
\label{barD2}
\eea
where 
\bea
 \cF_{\bar b ; \bar c \bar d} =  \cF_{\bar b ; \bar d \bar c}
 := \nabla_{\bar c} {\bar \o}_{ \bar b  \bar d } 
+ \nabla_{\bar d} {\bar \o}_{ \bar b  \bar c }  +4 {\bar \o}_{ \bar b  \bar e }\,
\G^{\bar e}_{ \bar c \bar d}~.
\eea 
Upon plugging the expression (\ref{barD2}) into (\ref{D2barD2}), there occur
several sectors that differ from each other by the number of superfields hit hy derivatives.
One particular sector, which is proportional to $\pa {\bar \F}   \pa {\bar \F} {\bar D}  {\bar \F}$, 
can be seen to vanish if and only if the equation (\ref{se2b}).
As a result, the above expression for $ \cF_{\bar b ; \bar c \bar d} $ simplifies
\bea
 \cF_{\bar b ; \bar c \bar d} =  4 {\bar \o}_{ \bar b  \bar e }\,
\G^{\bar e}_{ \bar c \bar d}~.
\eea 
If we now recall that $\pa_a \G^{\bar e}_{ \bar c \bar d} 
= R^{\bar e}{}_{ \bar c  a \bar d} $
is the  Riemann curvature, then  (\ref{D2barD2}) becomes 
\bea
\X_{\dt \a} [\f, \bar \f] 
 &=& \frac{1}{4} \int {\rm d}^4 x  \, D^\a\Big[ 
({\bar D}_{\dt \a}  {\bar \f}^{\bar b}) ({\bar D}_{\dt \b}  {\bar \f}^{\bar c}) 
({\bar D}^{\dt \b}  {\bar \f}^{\bar d} )\Big]
{\bar \o}_{\bar b \bar e} \,R^{\bar e}{}_{ \bar c  a \bar d} \,D_\a \F^a \non \\
&+& \frac{1}{8} \int {\rm d}^4 x  \, 
({\bar D}_{\dt \a}  {\bar \f}^{\bar b}) ({\bar D}_{\dt \b}  {\bar \f}^{\bar c}) 
({\bar D}^{\dt \b}  {\bar \f}^{\bar d} )\,D^\a\Big[
{\bar \o}_{\bar b \bar e} \, R^{\bar e}{}_{ \bar c  a \bar d} \,D_\a \F^a\Big]~. 
\eea
This expression vanishes due to the following two observations.
First,  it holds that 
\be
({\bar D}_{\dt \a}  {\bar \f}^{\bar b}) ({\bar D}_{\dt \b}  {\bar \f}^{\bar c}) 
({\bar D}^{\dt \b}  {\bar \f}^{\bar d} )
+({\bar D}_{\dt \a}  {\bar \f}^{\bar c}) ({\bar D}_{\dt \b}  {\bar \f}^{\bar d}) 
({\bar D}^{\dt \b}  {\bar \f}^{\bar b} )
+({\bar D}_{\dt \a}  {\bar \f}^{\bar d}) ({\bar D}_{\dt \b}  {\bar \f}^{\bar b}) 
({\bar D}^{\dt \b}  {\bar \f}^{\bar c} )=0~.
\ee
Second, the covariant constancy of ${\bar \o}_{\bar b \bar e}$ implies that 
the tensor ${\bar \o}_{\bar b \bar e} \, R^{\bar e}{}_{ \bar c  a \bar d} $
is completely symmetric in its ``barred'' indices,
\be
T_{a \bar b \bar c \bar d} := {\bar \o}_{\bar b \bar e} \, R^{\bar e}{}_{ \bar c  a \bar d} 
=T_{a (\bar b \bar c \bar d )} ~.
\ee
This completes the proof.

\small{

}

\end{document}